\documentclass[preprint,aps,eqsecnum,floatfix,showpacs]{revtex4}
\usepackage{dcolumn}
\usepackage{bm}
\usepackage{graphicx}
\begin{document}
\title{A relativistic calculation of the deuteron threshold electrodisintegration at backward angles}
\author{A.\ Arriaga}
\affiliation{Centro de F{\'\i}sica Nuclear da Universidade de Lisboa, 1649-003 Lisboa\\
Departamento de F{\'\i}sica, Faculdade de Ci\^encias da Universidade de Lisboa, 1700 Lisboa, Portugal} 
\author{R.\ Schiavilla}
\affiliation{Jefferson Lab, Newport News, VA 23606 \\
Department of Physics, Old Dominion University, Norfolk, VA 23529, USA}

\date{\today}

\begin{abstract}
The threshold electrodisintegration of
the deuteron at backward angles is studied with a relativistic Hamiltonian,
including a relativistic one-pion-exchange potential (OPEP) with
off-shell terms as predicted by pseudovector coupling of pions to nucleons.
The bound and scattering states are obtained in the center-of-mass frame,
and then boosted from it to the Breit frame, where the evaluation of
the relevant matrix elements of the electromagnetic current operator
is carried out.  The latter includes, in addition to one-body, also
two-body terms due to pion exchange, as obtained, consistently with
the OPEP, in pseudovector pion-nucleon coupling theory.  The full Lorentz
structure of these currents is retained.  In order to estimate the
magnitude of the relativistic effects we perform, for comparison,
the calculation with a non-relativistic phase-equivalent Hamiltonian
and the standard non-relativistic expressions for the one-body and
two-body pion-exchange currents.  Our results for the electrodisintegration
cross section show that, in the calculations using one-body currents,
relativistic corrections become significant ({\it i.e.}, larger
than 10\%) only at high momentum transfer $Q$ ($Q^2 \simeq 40$fm$^{-2}$
and beyond).  However, the inclusion of two-body currents makes the
relativistic predictions considerably smaller than the corresponding
non-relativistic results in the $Q^2$ region (18--40) fm$^{-2}$.  The
calculations based on the relativistic model also confirm the inadequacy,
already established in a non-relativistic context, of the present
electromagnetic current model to reproduce accurately the experimental
data at intermediate values of momentum transfers.
\end{abstract}


\maketitle
\section{Introduction}
\label{sec:intro}

The problem of how to treat the relativistic dynamics of interacting,
composite objects, such as nucleons, is highly non trivial, and a
variety of different approaches have been developed.  These fall
essentially into two classes: either field-theory inspired methods,
such as, for example, the spectator~\cite{Gross69} and
Blankenbecler-Sugar~\cite{Blankenbecler66} covariant reductions of
the Bethe-Salpeter equation, or methods based on relativistic
Hamiltonian dynamics (for a review, see Ref.~\cite{Keister91}).
The former include explicitly anti-particle degrees of freedom
and are manifestly covariant, while the latter subsume these degrees
of freedom into effective potentials, only retain particle ({\it i.e},
positive energy) propagation in the intermediate states, and typically satisfy
the requirements of relativistic covariance only approximately---these
and additional issues are discussed in considerable detail in a review
by Gilman and Gross~\cite{Gilman02}.

Both these methods---field-theory inspired and relativistic Hamiltonian
dynamics---have been used in calculations of few-nucleon properties,
including binding energies, momentum distributions, and electromagnetic
form factors.  Among the many references (a large, but non-exhaustive,
listing of them is in Ref.~\cite{Gilman02}), we only mention here the
calculations of: deuteron form factors~\cite{Vanorden95} and triton
binding energy~\cite{Stadler97} in the spectator-equation formalism;
deuteron form factors within the framework of relativistic Hamiltonian
dynamics, in the front-form~\cite{Lev00}, point-form~\cite{Allen01},
and instant-form~\cite{Schiavilla02} implementations of it; binding
energies and momentum distributions of $A$=3 and 4 nuclei in
instant-form Hamiltonian dynamics~\cite{Forest99,Carlson93}.

In the present work we study the deuteron threshold electrodisintegration
at backward angles with a relativistic Hamiltonian, including a relativistic
one-pion-exchange potential with off-energy-shell terms as predicted by
pseudovector coupling of pions to nucleons.  The electromagnetic current is
taken to consist of one- and two-body terms, the latter too derived from pseudovector
pion-nucleon interactions.  The full Lorentz structure of these currents is
retained in the calculation of their matrix elements between the initial deuteron
and final $n$$p$ continuum states.  Corrections associated with the boosting
of these states from the center-of-mass to the Breit frame, in which the
evaluation of the matrix elements is actually carried out, are also taken
into account.

The deuteron threshold electrodisintegration proceeds predominantly
via a magnetic-dipole transition between the bound deuteron and $^1$S$_0$
scattering state.  Since the early seventies, it has been known~\cite{Hockert73}
that the associated (isovector) transition form factor is dominated,
at momentum transfers of $\simeq$ (8--16) fm$^{-2}$, by the contributions
of two-body currents of pion range.  To the best of our knowledge, all
calculations of the cross section for this process have been carried out
so far within essentially a non-relativistic framework (see, for example,
Refs.~\cite{Leidemann90,Schiavilla91}).  One exception we are aware of
is the front-form Hamiltonian dynamics study of Ref.~\cite{Keister88},
which, however, only included single-nucleon currents. 

One of the goals of the present work is to assess the importance of
relativistic effects in the deuteron threshold electrodisintegration.
To this end, we also perform the calculation of the cross section
with a non-relativistic Hamiltonian, phase-equivalent to the relativistic
Hamiltonian described above, and the (standard) non-relativistic limits
of the one-body and two-body pion-exchange current operators.

This paper is organized into five sections.  In Sec.~\ref{sec:waves}
we discuss the relativistic Hamiltonian from which the bound and
scattering states are obtained, and the method used to boost these states
from the center-of-mass to an arbitrary frame.  In Sec.~\ref{sec:current}
we list the relativistic expressions adopted for the one-body and two-body
pion-exchange currents, while in Sec.~\ref{sec:calc} we illustrate the
momentum-space evaluation of the relevant matrix elements entering
the cross section of the deuteron threshold electrodisintegration.
Finally, in Sec.~\ref{sec:res} we present the results along
with a discussion and concluding remarks.  Details of the
calculation are relegated in the Appendices.

\section{The $n$$p$ bound and scattering wave functions}
\label{sec:waves}

The relativistic Hamiltonian used to generate the bound
and scattering wave functions in the $n$$p$ rest frame
is taken to be~\cite{Forest99,Carlson93,Forest00}
\begin{equation}
H^\mu = 2\, \sqrt{p^2 + m^2} + v^\mu  \ ,
\label{eq:ham}
\end{equation}
where $v^\mu$ consists of a short-range part $v_R$ parameterized
as in the Argonne $v_{18}$ potential~\cite{Wiringa95}, and of a
relativistic one-pion-exchange potential (OPEP) given by
\begin{eqnarray}
v^\mu_\pi ({\bf p}^{\, \prime},{\bf p})=&-&\frac{f_{\pi NN}^2}{m_\pi^2}
\frac{m}{E^{\, \prime}} \frac{f^2_\pi(k)}{ m_\pi^2+k^2} \frac{m}{E}
\Bigg[ {\bm \sigma}_1 \cdot {\bf k} \,
  {\bm \sigma}_2 \cdot {\bf k} \nonumber \\
&+& \mu \ (E^{\, \prime}-E)
\left( \frac{ {\bm \sigma}_1 \cdot {\bf p}^{\, \prime} \,
              {\bm \sigma}_2 \cdot {\bf p}^{\, \prime} } {E^{\, \prime}+m}
      -\frac{ {\bm \sigma}_1 \cdot {\bf p} \,
              {\bm \sigma}_2 \cdot {\bf p} } {E+m} \right) \Bigg]
{\bm \tau}_1 \cdot {\bm \tau}_2 \ . 
\label{eq:vpi}
\end{eqnarray}
Here $m$ denotes the nucleon mass, $f_{\pi NN}$ is the pion-nucleon
coupling constant ($f^2_{\pi NN}/4\pi$ =0.075), ${\bf p}$ and
${\bf p}^{\, \prime}$ are the initial and final relative momenta
in the center-of-mass frame, $E=\sqrt{p^2+m^2}$ and
$E^{\, \prime}=\sqrt{p^{\, \prime\, 2}+m^2}$ are the corresponding energies,
and ${\bf k}= {\bf p}-{\bf p}^{\, \prime}$ is the momentum transfer.
The monopole form factor $f_\pi(k)=(\Lambda_\pi^2 - m_\pi^2)/(\Lambda_\pi^2 +k^2)$
with $\Lambda_\pi=1.2$ GeV/c is considered in the present work.

The $\mu$-dependent term characterizes possible off-energy-shell
extensions of OPEP and, in particular, the value $\mu$=1 ($\mu=-1$)
is predicted by pseudovector (pseudoscalar) coupling of pions to
nucleons, while $\mu$=0 corresponds to the so-called \lq\lq minimal
non-locality\rq\rq choice~\cite{Friar77}.  As shown by Friar almost
three decades ago~\cite{Friar77}, these various off-shell extensions
of OPEP are related to each other by a unitary transformation, that is
\begin{equation}
H^\mu = {\rm e}^{-{\rm i}\mu U} H^{\mu=0} {\rm e}^{{\rm i}\mu U}
\simeq H^{\mu=0} + {\rm i}\, \mu\ \Big[ H^{\mu=0} \, , \, U \Big] \ ,
\label{eq:ut}
\end{equation}
if terms of $2\pi$-range (and shorter-range) are neglected.  The
hermitian operator $U$ is given explicitly in Ref.~\cite{Forest00}.
This unitary equivalence implies that predictions for electromagnetic
observables, such as the deuteron electrodisintegration cross section
under consideration here, are independent of the particular off-shell
extension adopted for OPEP, provided that the electromagnetic current
operator, specifically its two-body components associated with pion
exchange, is derived consistently with this off-shell extension.
As discussed later in Sec.~\ref{sec:current}, the pion-exchange
two-body currents used in this work have been obtained assuming
pseudovector coupling, and therefore the $\mu=1$ prescription is
taken for OPEP.  From now on, the $\mu=1$ superscript is dropped
from $H^\mu$ in Eq.~(\ref{eq:ham}) for simplicity.
The resulting relativistic Hamiltonian has been constructed to be
phase-equivalent to the non-relativistic $H$, based on the
Argonne $v_{18}$ potential.

The momentum-space wave functions of the deuteron and $n$$p$
scattering states are denoted respectively as $\psi_M({\bf p};0)$
and $\psi^{(-)}_{{\bf k};SM_S,T}({\bf p};0)$, where ${\bf p}$ is the
relative momentum and the zero in the argument
indicates the rest frame in which the deuteron and
$n$$p$ pair have velocity ${\bf V}$=0.  The bound-state wave function
with spin projection $M$ is written as in Ref.~\cite{Schiavilla02},
whereas the wave function corresponding to a scattering state with
the $n$$p$ pair having relative momentum ${\bf k}$, and spin, spin projection, and
isospin $S,M_S$, and $T$ ($M_T$=0 for $n$$p$), respectively, is
obtained from solving the Lippmann-Schwinger equation in momentum
space:
\begin{eqnarray}
\psi^{(-)}_{{\bf k};SM_S,T}({\bf p};0)&=&\phi_{{\bf k};SM_S,T}({\bf p};0)
\nonumber \\
&+&\sum_{M^\prime_S} \int\frac{d{\bf k}^\prime}{(2\pi)^3} \frac{1}{2}
\frac{1}{E_k -E_{k^\prime} - i\epsilon}
\left[T^{ST}_{M_S,M_S^\prime}({\bf k},{\bf k}^\prime)\right]^*
\phi_{{\bf k}^\prime;SM^\prime_S,T}({\bf p};0) \ ,
\label{eq:lse}
\end{eqnarray}
where the $\phi$'s are antisymmetric two-nucleon free states---hence
the factor 1/2 in the integral over intermediate states
${\bf k}^\prime$---with
\begin{equation}
\phi_{{\bf k};SM_S,T}({\bf p};0)=\frac{(2\pi)^3 }{\sqrt{2}}
\left[ \delta({\bf k}-{\bf p}) -(-)^{S+T}
\delta({\bf k}+{\bf p})\right] \chi^{S,T}_{M_S,0} \ ,
\label{eq:pw}
\end{equation}
and $E_k=2\, \sqrt{k^2+m^2}$ and similarly for $E_{k^\prime}$.
In Eq.~(\ref{eq:lse}) note that the $\psi^{(-)}_{{\bf k};SM_S,T}$'s
satisfy incoming-wave boundary conditions, since these are the
wave functions relevant for the process under consideration
here, and that they have been expressed in terms of the
$T$-matrix, defined as 
\begin{equation}
T^{ST}_{M_S,M_S^\prime}({\bf k},{\bf k}^\prime)
=\langle\psi^{(-)}_{{\bf k};SM_S,T}(0) \mid v \mid
        \phi_{{\bf k}^\prime;SM^\prime_S,T}(0)\rangle \ .
\end{equation}
In Eq.~(\ref{eq:pw}), $\chi^{S,T}_{M_S,0}$ denotes the $n$$p$
spin-isospin state $SM_S,TM_T=0$.

Bound or scattering wave functions in a frame moving with
velocity ${\bf V}$ with respect to the rest frame are obtained
from~\cite{Schiavilla02,Friar77}
\begin{equation}
\label{eq:boost}
\psi({\bf p};{\bf V})\equiv B({\bf p},{\bf V})\, \psi({\bf p}_\parallel/\gamma, {\bf p}_\perp;0) =
\frac{1}{\sqrt{\gamma}} \left[ 1 -\frac{{\rm i}}{4m} {\bf V}
\cdot ({\bm \sigma}_1-{\bm \sigma}_2)\times {\bf p} \right]
\psi({\bf p}_\parallel/\gamma, {\bf p}_\perp;0) \ ,
\end{equation}
where $\gamma = 1/\sqrt{1-V^2}$, and ${\bf p}_\parallel$
and ${\bf p}_\perp$ denote the components of the momentum
${\bf p}$ parallel and perpendicular to ${\bf V}$, respectively.
Only kinematical boost corrections are retained, in particular
the spin-dependent ones associated with Thomas precession 
are only included to order $V^2$.  The interaction-dependent
corrections are ignored.  However, it is interesting
to note that Eqs.~(\ref{eq:lse}) and~(\ref{eq:boost}) suggest that,
in order to boost the (fully interacting) scattering
state, one only needs to know how to boost the free states.

\section{Nuclear electromagnetic current}
\label{sec:current}

The electromagnetic current is taken as a sum of one-
and two-body terms 
\begin{equation}
{\bf j}= \sum_{i=1,2} {\bf j}_i({\bf p}_i^{\, \prime},{\bf p}_i)+
{\bf j}_{12}({\bf p}_1^{\, \prime},{\bf p}_2^{\, \prime},{\bf p}_1,{\bf p}_2) \ .
\label{eq:j1t}
\end{equation}
The one-body term corresponds to the space part of the
single-nucleon current $j_i^\alpha=(j_i^0,{\bf j}_i)$, with
\begin{equation}
j_i^\alpha({\bf p}_i^{\, \prime},{\bf p}_i)=
\bar{u}({\bf p}_i^{\, \prime})
\left[F_{1,i}(Q^2)\ \gamma^\alpha + \frac{i}{2m}\, F_{2,i}(Q^2)\ 
\sigma^{\alpha \beta} q_\beta \right] u({\bf p}_i) \ ,
\label{eq:j1r}
\end{equation}
where $u({\bf p}_i)$ and $\overline{u}({\bf p}_i^{\, \prime})$ ($\overline{u}
\equiv u^\dagger \gamma^0$) are the initial and final spinors of nucleon
$i$, $\sigma^{\alpha\beta}=(i/2)\left[\gamma^\alpha \, ,
\, \gamma^\beta\right]$, and $F_{1,i}(Q^2)$ and $F_{2,i}(Q^2)$
denote respectively the nucleon's Dirac and Pauli form factors, 
\begin{equation}
F_{a,i}(Q^2)\equiv \left[F_a^S(Q^2)+F_a^V(Q^2) \, \tau_{i,z}\right]/2 \ ,
\qquad a=1,2 \ .
\end{equation}
These form factors are normalized
as $F^S_1(0)$=$F^V_1(0)$=1 and $F^S_2(0)$=$-$0.12 n.m.~and $F^V_2(0)$=3.706
n.m.~(in units of nuclear magnetons).  The H\"ohler
parameterization~\cite{Hohler76} of $F_1$ and $F_2$ is used
in this work.  The spinor $u$, or rather its adjoint, is given by
\begin{equation}
u^\dagger({\bf p})=\left( \frac{E+m}{2 E}\right)^{1/2}
\left( \chi^\dagger_{\sigma\tau}\, , \,
\chi^\dagger_{\sigma\tau} \frac{ {\bm \sigma} \cdot
 {\bf p} }{E+m} \right) \ ,
\label{eq:ubsp}
\end{equation}
where ${\bf p}$ and $E$=$\sqrt{p^2+m^2}$ are
the nucleon's momentum and energy, and $\chi_{\sigma\tau}$ is its
(two-component) spin-isospin state.  Note that $u^\dagger u$=$1$.
Finally, the four-momentum transfer $q^\mu$, with $Q^2$=$-q^\mu q_\mu$,
is taken in the Breit frame, in which the initial deuteron has momentum
$-{\bf q}/2$ and the final $n$$p$ pair has momentum $+{\bf q}/2$,
and is given by $q^\mu=(\omega,q\hat{\bf z})$ with $\omega=E_f-E_i$,
where $E_i=\sqrt{m_d^2+q^2/4}$ ($m_d$ is the deuteron rest mass) and
$E_f=\sqrt{E_k^2+q^2/4}$ ($E_k$ is the center-of-mass energy of the
$n$$p$ pair, {\it i.e.} $E_k=2\, \sqrt{k^2+m^2}$).

Assuming pseudovector $\pi$-$N$ coupling, the two-body current
associated with pion exchange is written as
\begin{equation}
{\bf j}_{12}({\bf p}_1^{\, \prime},{\bf p}_2^{\, \prime},{\bf p}_1,{\bf p}_2)=
{\bf j}_{12}^{\,(a)}({\bf p}_1^{\, \prime},{\bf p}_2^{\, \prime},{\bf p}_1,{\bf p}_2)+
{\bf j}_{12}^{\,(b)}({\bf p}_1^{\, \prime},{\bf p}_2^{\, \prime},{\bf p}_1,{\bf p}_2) \ ,
\end{equation}
where ${\bf j}_{12}^{\, (a)}$ is the current corresponding to the two
seagull diagrams, and ${\bf j}_{12}^{\, (b)}$ is the current associated
with the pion in flight diagram.  They are given by 
\begin{eqnarray}
\label{eq:j12a}
{\bf j}_{12}^{\, (a)}\left({\bf p}^\prime_1,{\bf p}^\prime_2, {\bf p}_1, {\bf p}_2 \right)
&=&i\, G^V_E(Q^2)
\left({\mbox{\boldmath $\tau$}_1 \times \mbox{\boldmath $\tau$}_2} \right)_z
\frac{f^2_{\pi NN}}{m^2_\pi}
\frac{f^2_\pi(k_2)}{k_2^{\, 2}-k_2^{0\,2} + m_\pi^2} \nonumber \\
& & \bar {u}({\bf p}_1^{\, \prime}) \mbox{\boldmath $\gamma$} 
\gamma_5 u({\bf p}_1)\, 
\left[ k_2^\nu\, \bar {u}({\bf p}_2^{\, \prime}) \gamma_\nu \gamma_5 
u({\bf p}_2) \right] + 1\rightleftharpoons 2 \ , 
\end{eqnarray}
\begin{eqnarray}
\label{eq:j12c}
{\bf j}_{12}^{\, (b)} \left({\bf p}^\prime_1,{\bf p}^\prime_2, {\bf p}_1, {\bf p}_2 \right) 
&=& i\, G^V_E(Q^2)
\left({\mbox{\boldmath $\tau$}_1 \times \mbox{\boldmath $\tau$}_2} \right)_z
\frac {f^2_{\pi NN}}{m^2_\pi}
\frac{f^2_\pi(k_1)}{k_1^{\, 2}-k_1^{0\,2} + m_\pi^2}
\frac{f^2_\pi(k_2)}{k_2^{\, 2}-k_2^{0\,2} + m_\pi^2} \nonumber \\
& &\left( {\bf k}_1 - {\bf k}_2 \right) 
 \left[k_1^\nu\, \bar {u}({\bf p}_1^{\, \prime}) \gamma_\nu 
\gamma_5 u({\bf p}_1) \right] 
\left[k_2^\rho\, \bar {u}({\bf p}_2^{\, \prime}) \gamma_\rho \gamma_5 
u({\bf p}_2) \right] \ ,
\end{eqnarray}
where the four-momentum $k_i^\mu \equiv (k_i^0,{\bf k}_i)$, $i=1,2$, has
$k_i^0$=$E_i^\prime -E_i$ and ${\bf k}_i$=${\bf p}^\prime_i-{\bf p}_i$,
and $f_{\pi NN}$ and $f_\pi(k_i)$ are respectively the
pion-nucleon coupling constant and monopole form factor
introduced previously.  The fractional momenta ${\bf k}_1$ and
${\bf k}_2$ delivered to nucleons 1 and 2 add up to ${\bf q}$, that
is ${\bf k}_1+{\bf k}_2$=${\bf q}$.  The nucleon isovector Sachs form
factor $G^V_E(Q^2)$, related to $F_1^V(Q^2)$ and $F_2^V(Q^2)$ by
$G_E^V(Q^2)=F_1^V(Q^2)-(Q^2/4m^2)F_2^V(Q^2)$, is used in the
two-body currents.  This choice is motivated by the following
considerations.  In the non-relativistic limit, it is easy to
show that the two-body pion-exchange current satisfies current
conservation with the (non-relativistic) OPEP, obtained from
Eq.~(\ref{eq:vpi}) by setting $E$=$E^\prime$=$m$, if the same
electromagnetic form factor is used in the charge operator
and longitudinal component of the current.  As shown in
Appendix~\ref{a:cnt}, $G_E^V(Q^2)$ is used in the non-relativistic
expression of the charge operator.  Of course, the continuity
equation places no restrictions on the electromagnetic form factors
that may be used in the transverse components of the current.
Ignoring this ambiguity, the choice $G_E^V(Q^2)$ satisfies the
``minimal'' requirement for current conservation.  In the
relativistic case, we choose to keep this same electromagnetic
form factor in the two-body currents.

The full Lorentz structure of the one- and two-body currents
is retained in the calculations reported here.  The latter
are listed, along with their respective non-relativistic
limits, in Appendix~\ref{a:cnt} for completeness.

Finally, in earlier published work on the form factors
and threshold electrodisintegration cross section of the
deuteron~\cite{Wiringa95,Schiavilla91} and form factors of
the A=3--6 nuclei, most recently~\cite{Marcucci98,Wiringa98}, the
contributions associated with the boosts of the initial and final
wave functions were neglected, and only terms up to order $(v/c)^2$
were included in the non-relativistic expansion of $j_i^\alpha$, namely
the well known Darwin-Foldy and spin-orbit corrections to the charge
operator $j_i^0$.  Moreover, the two-body charge and current operators
were taken to leading order.

\section{Calculation}
\label{sec:calc}

In the one-photon-exchange approximation, the differential
cross section for deuteron electrodisintegration in the
laboratory frame can be expressed as~\cite{deForest66}
\begin{equation}
\label{eq:xsec}
\frac {d^2 \sigma}{d\varepsilon^\prime d\Omega^\prime}=\sigma_M 
\left[ W_2(Q^2,q_\mu P^\mu) +
W_1(Q^2,q_\mu P^\mu) \, {\rm tan}^2(\theta/2) \right]
\end{equation}
where $\varepsilon^\prime$ and $\Omega^\prime$ are the final
electron energy and solid angle, $\sigma_M$ is the Mott cross
section, and the invariant response functions $W_1$ and $W_2$
depend on the square of the four momentum transfer, denoted as
before by $Q^2$, and the Lorentz scalar $q^\mu P_\mu$, with $P_\mu$
being the four momentum of the deuteron in the initial state.
At backward angles, the cross section above is dominated by
$W_1$, {\it i.e.} transverse scattering. (Measurements
of the deuteron threshold electrodisintegration have been
performed at angles typically $\ge 155$$^{\circ}$,
see Sec.~\ref{sec:res}, for which ${\rm tan}^2(\theta/2)$ is
$\ge 20$.)  Hence, in the following, we will consider only
the response $W_1$.  In the Breit frame, defined in
Sec.~\ref{sec:current}, it can be written as 
\begin{equation}
W_1(q,\omega) =\sqrt{1+q^2/(2\,m_d)^2} \sum_{S,T=0,1} R^{ST}(q,\omega) \ ,
\end{equation}
where the contribution from the individual spin-isospin states of the
final $n$$p$ pair is given by
\begin{equation}
\label{eq:rft}
R^{ST}(q,\omega )=\frac {1}{3}\sum_{M, M_S} \int \frac {d{\bf k}}{(2\pi)^3}
\frac{1}{2} \left| {\bf A} (q \hat{\bf z},{\bf k};S,M_S,T,M) \right|^2 
\delta( E_i +\omega-E_f)  \ .
\end{equation}
In the equation above, as in Sec.~\ref{sec:waves}, $M$ is the
spin projection of the deuteron, and $S,M_S$, and $T$ specify
the spin, spin projection, and isospin of the $n$$p$ scattering
state, while in the energy-conserving $\delta$-function $E_i$
and $E_f$ are, respectively, the initial deuteron and final
$n$$p$-pair Breit-frame energies, $E_i=\sqrt{q^2/4+m_d^2}$ and
$E_f=\sqrt{q^2/4+E_k^2}$, with $E_k=2\,\sqrt{k^2+m^2}$.  Finally,
the three-momentum transfer ${\bf q}$ is taken along the
$\hat{\bf z}$ direction.

The amplitude ${\bf A}$ denotes the matrix elements of 
the transverse components ({\it i.e.}, orthogonal to ${\bf q}$)
of the current operator, namely 
\begin{equation}
{\bf A}(q\hat{\bf z},{\bf k};S,M_S,T,M)= 
\langle \psi^{(-)}_{{\bf k};SM_S,T}({\bf V}_f) | {\bf j}_\perp(q \hat{\bf z})
 |\psi_{M}({\bf V}_i)\rangle \ .
\label {eq:amp}
\end{equation}
Here, $\psi_{M}({\bf V}_i)$ and $\psi^{(-)}_{{\bf k};SM_S,T}({\bf V}_f)$
are the deuteron and $np$ scattering states boosted from the center-of-mass
frame, where they are calculated in momentum space with the methods discussed
in Sec.~\ref{sec:waves}, to the Breit frame, in which they have velocities
given by, respectively, ${\bf V}_i=-(q/2)\hat{\bf z}/E_i$ and
${\bf V}_f=+(q/2)\hat{\bf z}/E_f$.

By inserting Eq.~(\ref{eq:lse}) into in Eq.~(\ref{eq:amp}),
the amplitude ${\bf A}$ can be conveniently decomposed
into the sum of two terms:
\begin{equation}
\label {eq:ampl}
{\bf A}= {\bf A}^{\rm PW} +{\bf A}^{\rm FSI} \ ,
\end{equation}
where
\begin{equation}
\label{eq:amp-pw}
{\bf A}^{\rm PW}(q\hat{\bf z},{\bf k};S,M_S,T,M)=
\langle \phi _{{\bf k};SM_S,T}({\bf V}_f)|
{\bf j}_\perp(q\hat{\bf z})|\psi_M({\bf V}_i)\rangle \ ,
\end{equation}
and
\begin{equation}
\label{eq:amp-fsi}
{\bf A}^{\rm FSI}(q\hat{\bf z},{\bf k};S,M_S,T,M)
=\sum_{M_S^\prime} \int\frac {d{\bf k}^\prime}{2\,(2\pi)^3}
\frac{T^{ST}_{M_S,M^\prime_S}({\bf k},{\bf k}^\prime)}
{E_k - E_{k^\prime} + i\epsilon}
{\bf A}^{\rm PW}(q\hat{\bf z},{\bf k}^\prime;S,M_S^\prime,T,M) \ .
\end{equation}
Thus, the amplitude $A^{\rm PW}$ corresponds to describing 
the final $n$$p$ states by plane waves (PW), while the amplitude
$A^{\rm FSI}$ takes into account interaction effects in these states.

The electromagnetic current operator includes the one- and two-body
terms discussed in the previous section.  Details of the calculation
of the amplitudes are reported in Appendix~\ref{sec:amplitudes}.

\section{Results and Conclusions}
\label{sec:res}

In this section we report the results obtained in the
laboratory frame for the cross section of the deuteron
threshold electrodisintegration at backward angles.  The
calculations were carried out with the relativistic (R)
Hamiltonian of Sec.~\ref{sec:waves}, including the OPEP
with off-energy-shell extension predicted by pseudovector
coupling of pions to nucleons, {\it i.e.} with $\mu$=+1 in
Eq.~(\ref{eq:vpi}).  This Hamiltonian was constructed to be
phase-equivalent to the non-relativistic (NR) Hamiltonian,
based on the Argonne $v_{18}$ potential~\cite{Wiringa95},
using the methods developed in Ref.~\cite{Carlson93}.

\begin{figure}[bth]
\includegraphics[angle=-90,width=6in]{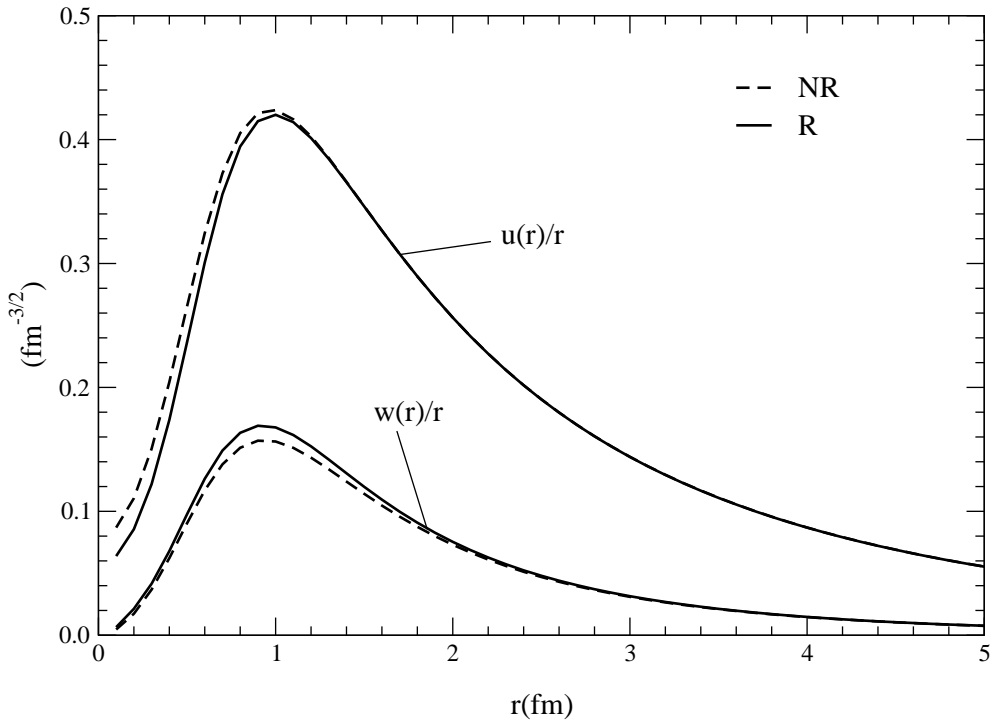}
\caption{The deuteron S- and D-state radial wave functions
obtained with the R and NR Hamiltonian models.}
\label{fig:fig1}
\end{figure}

\begin{figure}[nth]
\includegraphics[angle=-90,width=6in]{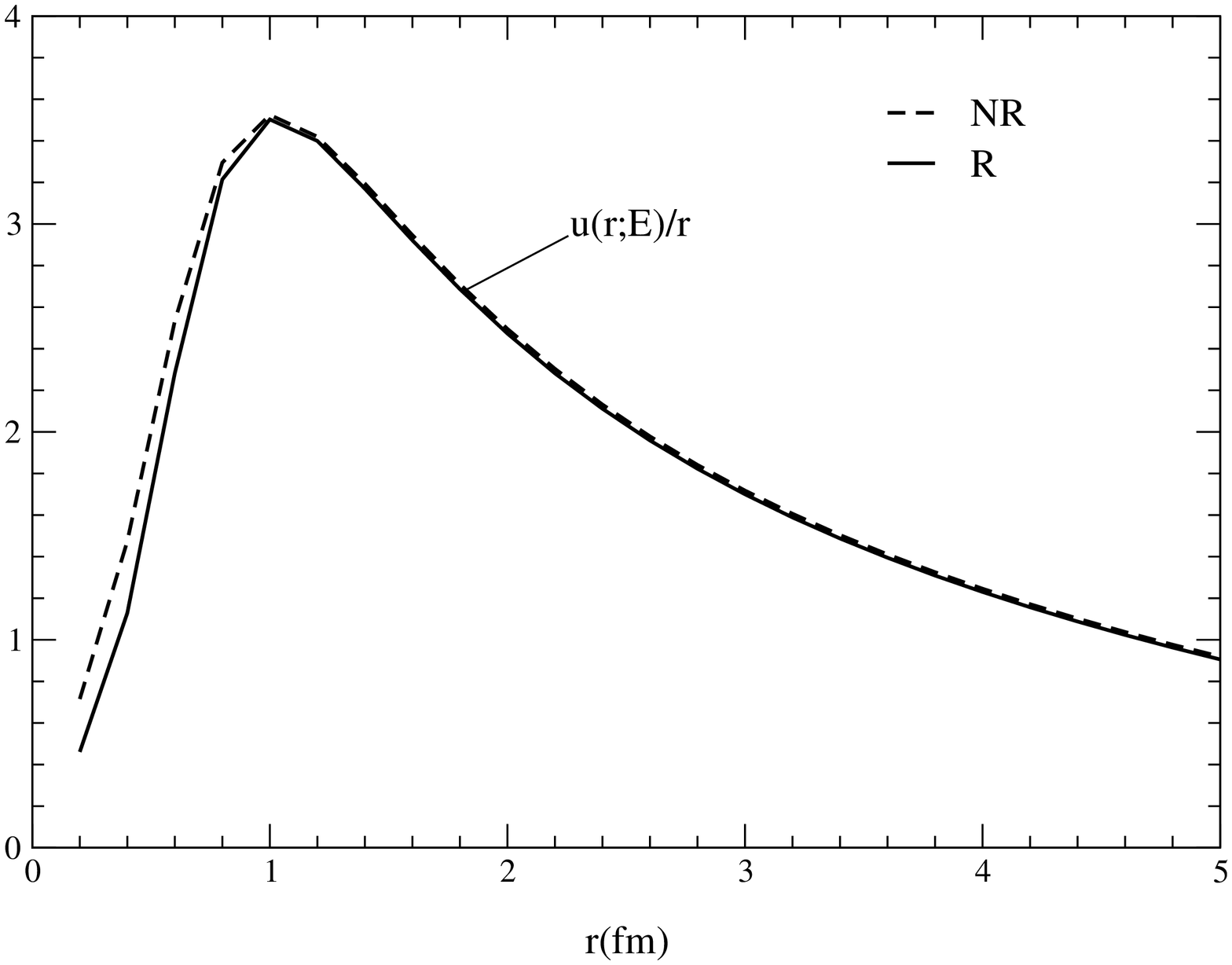}
\caption{The $n$$p$ $^1$S$_0$-state
radial wave functions obtained with the R and NR Hamiltonian
models at a center-of-mass energy of 1.5 MeV}
\label{fig:fig2}
\end{figure}

In Figs.~\ref{fig:fig1} and~\ref{fig:fig2} we show
the deuteron and $n$$p$ $^1$S$_0$ wave functions,
respectively, derived from the R and NR Hamiltonian models
(the continuum wave function is calculated at a
center-of-mass energy of 1.5 MeV).  The deuteron R D-wave is
larger than the NR at inter-nucleon separations less than
1.5 fm, the corresponding D-state probabilities are
6.26\% (R) and 5.76\% (NR)---the difference has its
origin in the local and non-local characters of the NR
and R ($\mu$+1) OPEP (for a discussion of this point,
see Ref.~\cite{Forest00}).  However, the D- to S-state
ratio and quadrupole moment, which are more sensitive 
to the wave functions in the asymptotic region, are
respectively 0.0260 and 0.272 fm$^2$ in the R model,
and 0.0250 and 0.270 fm$^2$ in the NR~\cite{Forest00}. 
In contrast, the R and NR $^1$S$_0$ continuum
wave functions hardly differ from each other, due to
the vanishing of the tensor force in this channel.


\begin{figure}[bth]
\includegraphics[angle=-90,width=6in]{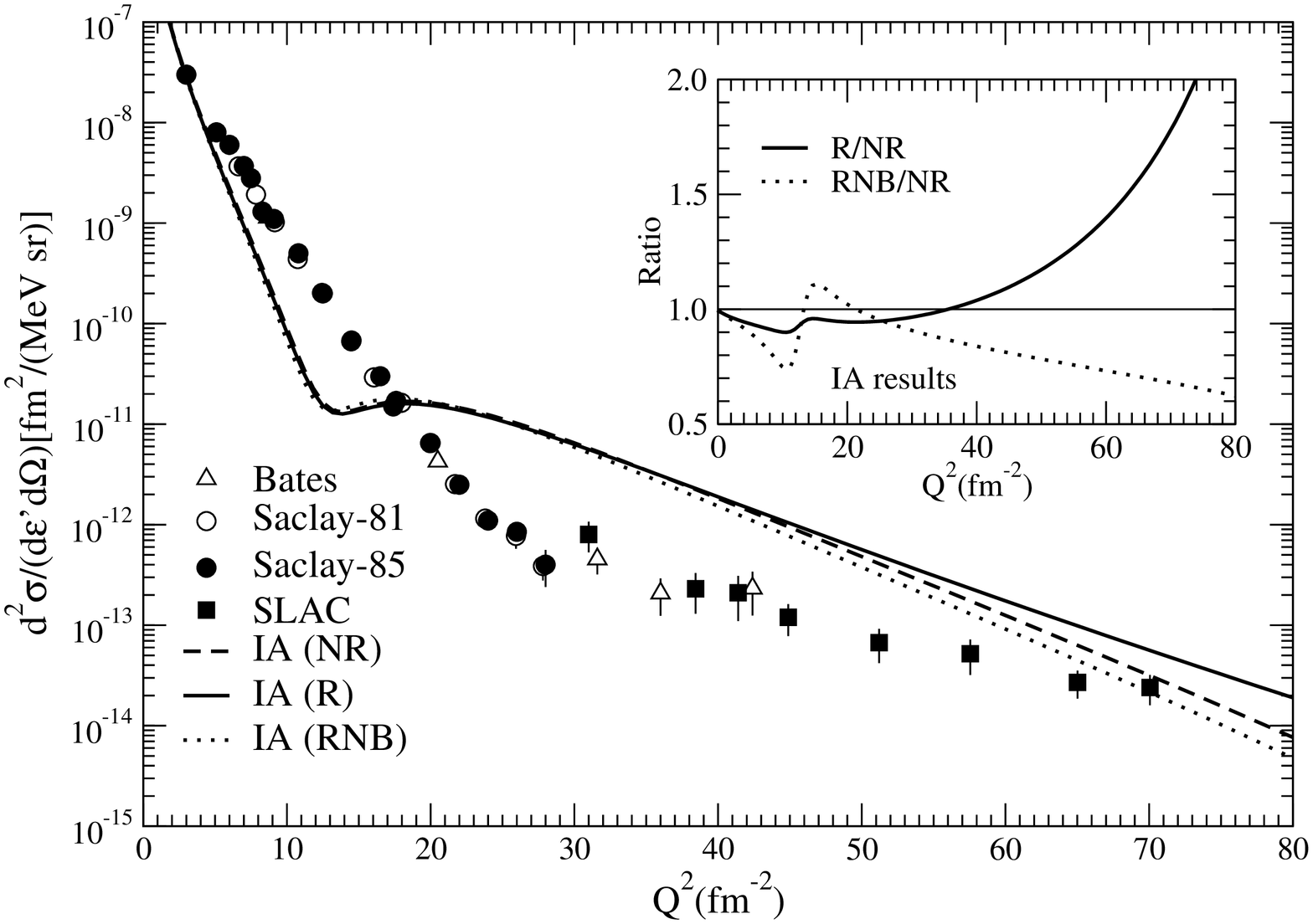}
\caption{The cross sections for deuteron threshold
electrodisintegration at backward angles, obtained with the
relativistic and non-relativistic Hamiltonian models
and corresponding one-body currents [curves labeled IA (R)
and IA (NR)] and by ignoring boost corrections
in the relativistic calculation [curve labeled
IA (RNB)], are compared with the experimental data from 
Refs.~\protect\cite{Schmitt97,Bernheim81,Auffret85,Frodyma93}.
The inset displays the ratio of the IA (R)
and IA (RNB) to the NR predictions.}
\label{fig:fig3}
\end{figure}

In Fig.~\ref{fig:fig3} we report the results for the
electrodisintegration cross section obtained in the
laboratory frame with the R (solid line) and NR (dashed
line) Hamiltonian models and corresponding one-body
currents, given respectively in Eqs.~(\ref{eq:j1})
and~(\ref{eq:j1nr}).  The cross section obtained
by ignoring the boost corrections for both the initial
and final states in the R calculation---this is equivalent
to setting ${\bf V}$=0 in Eq.~(\ref{eq:boost})---is
displayed by the dotted line, labeled RNB.
All calculated cross sections include the contributions
of $n$$p$ final states with total angular momentum
up to $J$=3.  These contributions are responsible for
filling in the well-known nodal structure at $Q^2 \simeq 12$
fm$^{-2}$ in the cross section obtained with one-body
currents by retaining only the $^1$S$_0$ channel in
the final state~\cite{Hockert73,Leidemann90,Schiavilla91,Keister88}
(see Fig.~\ref{fig:fig6} below).  Finally, the inset
of Fig.~\ref{fig:fig3} shows the ratios of the R and
RNB to the NR predictions.

The experimental data, labeled Bates, Saclay-81 and Saclay-85, are
respectively from Refs.~\cite{Schmitt97,Bernheim81,Auffret85},
and have been averaged over the interval 0--3 MeV of the recoiling 
$n$$p$ pair center-of-mass energy; those labeled SLAC are
from Ref.~\cite{Frodyma93}, and have been averaged over
the interval 0--10 MeV.  However, all theoretical curves 
in this figure, and following ones, have been calculated at
a fixed center-of-mass energy of 1.5 MeV and at an electron
scattering angle $\theta$=155$^\circ$.  The effect of
the width of the energy interval above threshold of the
final state, over which the cross section values are
averaged, was studied in Ref.~\cite{Schiavilla91}, and
found to be very small.  The electron scattering angles
in the Saclay, Bates, and SLAC measurements were respectively
155$^\circ$, 160$^\circ$, and 180$^\circ$, but in fact the
calculated cross section is weakly dependent on the
specific value of the backward angle, 
since $\sigma_M \, {\rm tan}^2(\theta/2) \rightarrow 
\alpha^2/(4\, \epsilon^2)$ as $\theta \rightarrow 180^\circ$
(here, $\alpha$ is fine structure constant and $\epsilon$
the initial electron energy).

The inset of Fig.~\ref{fig:fig3} shows that the IA (R)
and IA (NR) predictions differ significantly ({\it i.e.},
more than 10\%) only for $Q^2 > 45$ fm$^{-2}$.  At lower
momentum transfers, the IA (R) cross section values are
within 10\% of the IA (NR).  Comparison of the IA (R) and
IA (RNB) results shows the effect of the boosts
corrections in the initial deuteron and final $n$$p$ states.
We have verified explicitly, by switching off the Thomas
precession term in the boost operator of Eq.~(\ref{eq:boost}),
that the dominant correction arises from the Lorentz
contraction term (the resulting curve is essentially
indistinguishable from that labeled IA (R); it is not
shown to reduce clutter).  Indeed, the IA (RNB) results can
be approximately overlaid over the IA (R) results by multiplying
the former by the factor $[1+Q^2/(16\,m^2)]$, corresponding
to the square of the Lorentz factor $\gamma \simeq \gamma_i 
\simeq \gamma_f =1/\sqrt{1-V^2}$,
where $V =|{\bf V}| = (q/2)/\sqrt{4\, m^2+q^2/4}\,$, and the
deuteron binding energy and $n$$p$ center-of-mass energy
have been neglected.  A similar effect was discussed in
Ref.~\cite{Schiavilla02} in the context of a calculation
of the deuteron electromagnetic form factors: it conforms with
the naive expectation that the overlap between the initial
and final states in configuration space is ``squeezed'' in the
direction of motion (namely, along ${\bf q}$) by $\gamma$ or,
equivelently, that its momentum space overlap is ``pushed out''
by $\gamma$.

\begin{figure}[bth]
\includegraphics[angle=-90,width=6in]{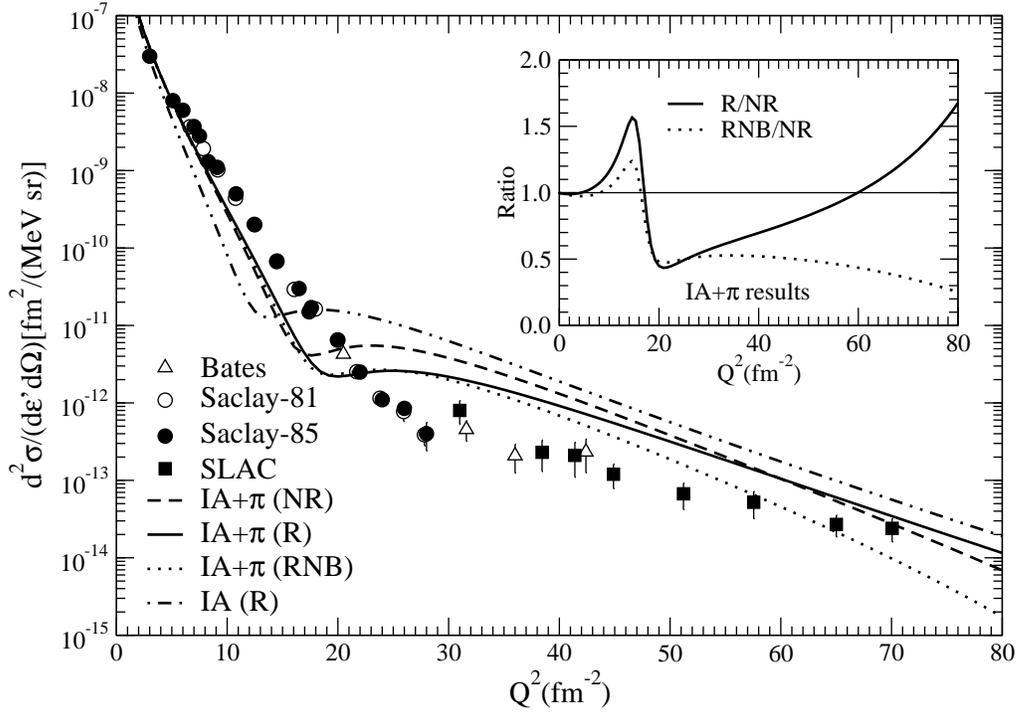}
\caption{Same as in Fig.~\protect\ref{fig:fig3}, but with
one-body and pion-exchange two-body currents.  The cross sections
obtained in the IA (R) calculation are also shown.}
\label{fig:fig4}
\end{figure}
In Fig.~\ref{fig:fig4} we report the cross section results
obtained by including, in addition to the single-nucleon current,
the two-body current associated with pion exchange (curves
labeled IA+$\pi$ with NR, R, and RNB), while in the inset we
display the ratios of R and RNB to NR predictions.  For reference,
we also show the IA (R) cross sections presented in Fig.~\ref{fig:fig3}.
The cross section values in the R calculation are significantly
smaller than those in the NR in the momentum transfer range
$Q^2$=18--40 fm$^{-2}$.  In fact, close inspection of
Figs.~\ref{fig:fig3}--\ref{fig:fig4} shows that
the pion exchange contribution in the R calculation
is larger than in the NR.  In both of these calculations, this
contribution is found to have the same sign, for $Q^2$
up to $\approx 12$ fm$^{-2}$, as the one-body contribution.  At larger
$Q^2$ values, however, the latter changes sign, and the
resulting destructive interference between it and the
two-body contribution is responsible for the suppression 
of the R cross section relative to the NR in this $Q^2$ region.

\begin{figure}[bth]
\includegraphics[angle=-90,width=6in]{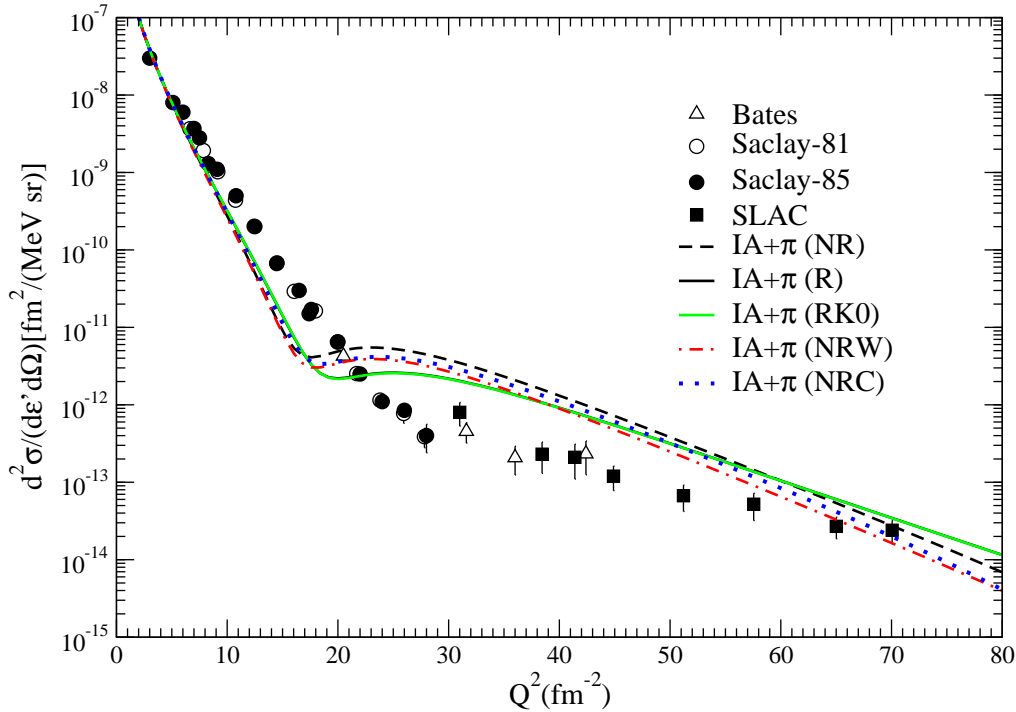}
\caption{(Color online) The IA+$\pi$ (R) and IA+$\pi$ (NR) predictions of Fig.~\protect\ref{fig:fig4} 
are compared with the results
corresponding to different approximations, labeled respectively
IA+$\pi$ (RK0), IA+$\pi$ (NRW), and IA+$\pi$ (NRC).  See text
for discussion.  The experimental data are from Refs.~\protect\cite{Schmitt97,Bernheim81,
Auffret85,Frodyma93}.}
\label{fig:fig5}
\end{figure}

To investigate the mechanisms responsible for the suppression
of the R relative to NR predictions for $Q^2$ in
the range 18--40 fm$^{-2}$, we have carried out two different
calculations, the results of which are displayed in Fig.~\ref{fig:fig5}.
In the first, labeled IA+$\pi$ (NRW), we have replaced the NR
expressions for the one- and two-body currents with the corresponding R ones,
in order to isolate relativistic effects in the currents.  Comparison between
the IA+$\pi$ (NRW) and IA+$\pi$ (NR) curves shows that these effects 
reduce the cross section, for $Q^2>18$ fm$^{-2}$.

In the second calculation, labeled IA+$\pi$ (NRC), we have used
NR one- and two-body currents but R deuteron and
$n$$p$ scattering wave functions without boost corrections---so 
this is the same as IA+$\pi$ (NR) calculation but for the
replacement of the NR wave functions by the corresponding R ones---with
the objective of isolating relativistic effects generated by the Hamiltonian.
As in the previous case, we find that these reduce the cross section.

In Fig.~\ref{fig:fig5} we also show the results of an R calculation
in which the time components ($k_i^0$) of the exchanged pion four-momenta 
in both the vertex operators and propagators of the two-body
currents in Eqs.~(\ref{eq:j12al})--(\ref{eq:j12cl}) are set to zero,
curve labeled IA+$\pi$ (RK0).  The latter essentially overlaps
the IA+$\pi$ (R) curve.  We have also verified by direct calculation
that ignoring the retardation effects only in the pion propagators
again hardly changes the IA+$\pi$ (R) predictions.
Thus, the explicit energy dependence of the vertex operators
implied by pseudovector coupling of pions to nucleons has
a negligible effect.

\begin{figure}[bth]
\includegraphics[angle=-90,width=6in]{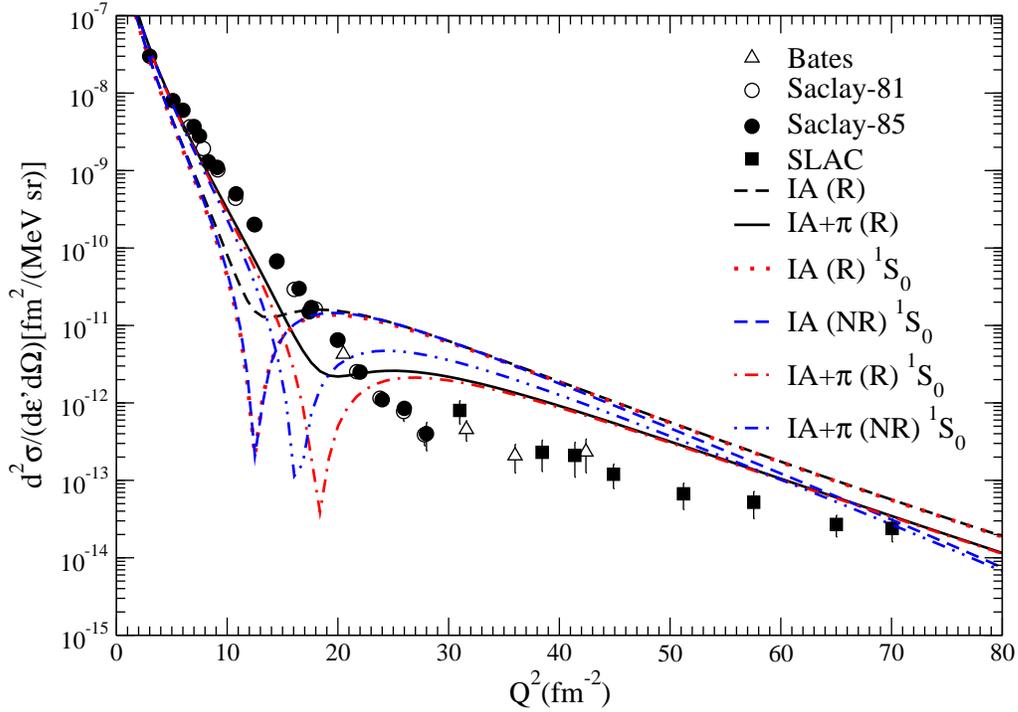}
\caption{(Color online)The cross sections for deuteron threshold
electrodisintegration at backward angles, obtained
in the IA (R) and IA+$\pi$ (R) calculations, are compared
with the results of calculations based on the NR and R Hamiltonian
models and corresponding one- and two-body currents, but including
only the $^1$S$_0$ channel in the $n$$p$ final state.  The experimental
data are from Refs.~\protect\cite{Schmitt97,Bernheim81,Auffret85,Frodyma93}.}
\label{fig:fig6}
\end{figure}
Figure~\ref{fig:fig6} shows the results of R and NR calculations
including only the $^1$S$_0$ channel in the $n$$p$ final state
compared both to data and the R results with the ``complete''
$n$$p$ state (all channels up to $J$=3), {\it i.e} the curves
labeled IA (R) and IA+$\pi$ (R) in Fig.~\ref{fig:fig4}.  The nodes
at $Q^2\approx 12$ fm$^{-2}$ (IA) and 18 and 16 fm$^{-2}$ (IA+$\pi$)
in the R and NR predictions including only the $^1$S$_0$ channel
are filled in by the contributions of higher partial waves in
the complete calculations.  The $^1$S$_0$ IA (R) and IA (NR)
cross sections are very close to each other, and thus confirm
the conclusions of Ref.~\cite{Keister88}, in which the
electrodisintegration cross sections was calculated within
a relativistic approach based on light-front-form Hamiltonian
dynamics, including only one-body currents.

To conclude, we find that relativistic effects in the calculations
with only one-body currents become important (larger than 10\%) at
momentum transfers $Q^2$ exceeding 40 fm$^{-2}$, and are
due, for the most part, to boost corrections.  However, when the
pion-exchange current contributions are also taken into account,
significant differences at lower $Q^2$ are obtained between the
cross sections predicted within the R and NR models for the Hamiltonian
and currents.  The interplay between 
relativistic effects in the interactions and currents conspire to
significantly reduce the cross section obtained in the R calculation
in the $Q^2$ range $\simeq 18-40$ fm$^{-2}$.

The cross section predictions based on both the R and NR models
do not reproduce the experimental data at $Q^2 > 10$ fm$^{-2}$,
thus demonstrating the inadequacy of the present model for
the electromagnetic current operator.  This conclusion corroborates
that of an earlier (NR) study~\cite{Schiavilla91}, and suggests
the need for including additonal (short-range) two-body currents.

Finally, in the present work we have not addressed the issue of
current conservation within the R framework.  Its discussion
would require constructing the two-body charge operator
associated with pion exchange, and studying the relation
between the pion-exchange charge and current operators and the
off-energy shell behavior of OPEP.  This is beyond the scope
of the present work.

\section*{Acknowledgments}
The work of A.A.~was partially supported by the Portuguese FCT
under contract POCTI/FNU/49505/2002, while that of R.S.~by the
U.S.~Department of Energy, Office of Nuclear Physics, under
contract DE-AC05-06OR23177.  One of the authors (A.A.)
would like to thank the JLab Theory Group for the support and warm
hospitality extended to her on several occasions.  Finally, some of
the calculations were made possible by grants of computing time from
the National Energy Research Supercomputer Center.
\appendix
\section{Current operator expressions}
\label{a:cnt}

In this Appendix we list the expressions for the one-body
and two-body pion-exchange current operators.  The time
(charge) and space (current) components of the one-body 
four-current read, respectively:
\begin{eqnarray}
\rho_i({\bf p}_i^{\,\prime},{\bf p}_i)=N_i^{\, \prime} N_i \Bigg[
F_{1,i}&+&F_{1,i}\, \frac{ {\bf p}_i^{\, \prime}  \cdot {\bf p}_i +
i {\bm \sigma}_i \cdot {\bf p}^{\, \prime}_i \times {\bf p}_i }
{(E_i^{\, \prime}+m)(E_i +m)} \nonumber \\
&+& \frac{ F_{2,i} }{2m} \,
 \left( \frac{ {\bf q} \cdot {\bf p}_i -
i {\bm \sigma}_i \cdot {\bf p}_i \times {\bf q} } {E_i + m}
      - \frac{ {\bf q} \cdot {\bf p}_i^{\, \prime} +
i {\bm \sigma}_i \cdot {\bf p}_i^{\, \prime} \times {\bf q} }
{E_i^{\, \prime} + m} \right) \Bigg] \ ,
\label{eq:rho1}
\end{eqnarray}
\begin{eqnarray}
{\bf j}_i({\bf p}_i^{\,\prime},{\bf p}_i) & & = N^{\, \prime}_i N_i\Bigg\{
F_{1,i}\ \left( \frac { {\bf p}_i - i {\bm \sigma}_i \times {\bf p}_i }{E_i+m}
+ \frac { {\bf p}^{\, \prime}_i + i {\bm \sigma}_i \times {\bf p}^{\, \prime}_i }
{E^{\, \prime}_i+m} \right) \nonumber \\
& & + \frac {F_{2,i}}{2m}\, \omega \,
\left( \frac { {\bf p}_i - i {\bm \sigma}_i\times {\bf p}_i }{E_i+m}
- \frac { {\bf p}^{\, \prime}_i + i 
{\bm \sigma}_i \times {\bf p}^{\, \prime}_i }{E^{\, \prime}_i+m} \right) \nonumber \\
& &-i \frac {F_{2,i}}{2m}\ {\bf q} \times {\bm \sigma}_i
\left[ 1+\frac{ {\bf p}_i^{\, \prime}\cdot{\bf p}_i}
{(E_i^{\, \prime} + m) (E_i + m)}\right]  +
i \frac {F_{2,i}}{2m}\
\frac{ {\bm \sigma}_i \cdot {\bf p}_i^{\, \prime} \
{\bf q} \times {\bf p}_i +
{\bm \sigma}_i \cdot {\bf p}_i\ {\bf q} \times {\bf p}_i^{\, \prime} }
{(E_i^{\, \prime} + m) (E_i + m)} \nonumber \\
& &+ \frac{F_{2,i}}{2m}\,
\frac{ {\bf q} \cdot {\bf p}_i \, {\bf p}_i^{\, \prime}\ -
{\bf q} \cdot {\bf p}^{\, \prime}_i \, {\bf p}_i }
{(E_i^{\, \prime} + m) (E_i + m)} \Bigg\} \ ,
\label{eq:j1}
\end{eqnarray}
where the spinor-normalization factor $N_i=\sqrt{(E_i+m)/(2\, E_i)}$
and similarly for $N_i^{\, \prime}$, ${\bm \sigma}_i$ is the Pauli
spin operator of nucleon $i$, and the initial and final nucleon spin-isospin states
$\chi_{\sigma_i^\prime\tau_i^\prime}$ and $\chi_{\sigma_i\tau_i}$ are not
explicitly shown.  The initial and final momenta
are denoted respectively as ${\bf p}_i$ and ${\bf p}_i^{\,\prime}$,
while the energy and three-momentum transfers $\omega$ and ${\bf q}$
are taken in the Breit frame, defined in Sec.~\ref{sec:current} after
Eq.~(\ref{eq:ubsp}).  The non-relativistic limits to order $(p/m)^2$
included are written as
\begin{eqnarray}
\rho^{\rm NR}_i({\bf p}_i^{\,\prime},{\bf p}_i)&=&
\frac{ G_{E,i} }{ \sqrt{1+Q^2/(4m)^2} }+i\frac{ 2\, G_{M,i}- G_{E,i} }{4 m^2}\
{\bm \sigma}_i \cdot {\bf p}^{\, \prime}_i\times {\bf p}_i \ ,
\label{eq:rho1nr} \\
{\bf j}^{\rm NR}_i({\bf p}_i^{\,\prime},{\bf p}_i)&=&
\frac{ G_{E,i} }{2  m}\left({\bf p}_i + {\bf p}_i^{\, \prime}\right)
+i \frac{ G_{M,i} }{2 m} \ {\bm \sigma}_i \times {\bf q} \ ,
\label{eq:j1nr}
\end{eqnarray}
where $Q^2=-q^\mu q_\mu$ and the Sachs nucleon form factors,
defined as
\begin{eqnarray}
G_{E,i}&=& F_{1,i} - \frac{Q^2}{4m^2}\, F_{2,i} \ , \\
G_{M,i}&=& F_{1,i} + F_{2,i} \ ,
\end{eqnarray}
have been introduced in Eqs.~(\ref{eq:rho1nr})--(\ref{eq:j1nr}).
The two-body pion-exchange currents are given by
\begin{eqnarray}
\label{eq:j12al}
{\bf j}_{12}^{\, (a)}\left({\bf p}^\prime_1,{\bf p}^\prime_2, {\bf p}_1, {\bf p}_2 \right)
&=&i\, N_1^\prime N_2^\prime N_1 N_2\, G^V_E(Q^2) \,
\frac{f^2_{\pi NN}}{m^2_\pi}\, \frac{f^2_\pi(k_2)}{k_2^{\, 2}-k_2^{0\,2} + m_\pi^2} \,
\left({\bm \tau}_1 \times {\bm \tau}_2\right)_z \nonumber \\
& & \left[ {\bm \sigma}_1 + \frac{ ({\bm \sigma}_1 \cdot {\bf p}^\prime_1) \,\, 
{\bm \sigma}_1\,\, ({\bm \sigma}_1 \cdot{\bf p}_1)}{(E^\prime _1+m)(E_1+m)} \right] 
\Bigg\{ k_2^0 \left( \frac{ {\bm \sigma}_2 \cdot {\bf p}^\prime_2}{E^\prime_2+m} +
\frac{ {\bm \sigma}_2 \cdot {\bf p}_2 }{E_2 + m} \right) \nonumber \\
&-&\left[ {\bm \sigma}_2\cdot{\bf k}_2 + \frac{ ({\bm \sigma}_2 \cdot {\bf p}^\prime_2) \,\,
{\bm \sigma}_2\cdot{\bf k}_2\,\, ({\bm \sigma}_2 \cdot{\bf p}_2)} {(E^\prime_2+m)(E_2+m)}
\right] \Bigg\} + 1\rightleftharpoons 2   \ ,
\end{eqnarray}
\begin{eqnarray}
\label{eq:j12cl}
{\bf j}_{12}^{\, (b)} \left({\bf p}^\prime_1,{\bf p}^\prime_2, {\bf p}_1, {\bf p}_2 \right)
\!\!&=&\!\!i\,N_1^\prime N_2^\prime N_1 N_2\, G^V_E(Q^2) \,
\frac {f^2_{\pi NN}}{m^2_\pi}
\frac{f^2_\pi(k_1)}{k_1^{\, 2}-k_1^{0\,2} + m_\pi^2}
\frac{f^2_\pi(k_2)}{k_2^{\, 2}-k_2^{0\,2} + m_\pi^2} 
\left({\bm \tau}_1 \times {\bm \tau}_2\right)_z \nonumber \\
({\bf k}_1-{\bf k}_2)\!\!&&\!\!\!\!\!\!\!\!
 \Bigg\{ k_1^0 \left( \frac{ {\bm \sigma}_1 \cdot {\bf p}^\prime_1}{E^\prime_1+m}\!+\!
\frac{ {\bm \sigma}_1 \cdot {\bf p}_1 }{E_1 + m} \right)
\!\!-\!\!\left[ {\bm \sigma}_1\cdot{\bf k}_1\!+\!  \frac{ ({\bm \sigma}_1 \cdot {\bf p}^\prime_1) \,\,
{\bm \sigma}_1\cdot{\bf k}_1\,\, ({\bm \sigma}_1 \cdot{\bf p}_1)} {(E^\prime_1+m)(E_1+m)}
\right] \Bigg\} \nonumber \\
&&\!\!\!\!\!\!\!\! \Bigg\{ k_2^0 \left( \frac{ {\bm \sigma}_2 \cdot {\bf p}^\prime_2}{E^\prime_2+m}\!+\!
\frac{ {\bm \sigma}_2 \cdot {\bf p}_2 }{E_2 + m} \right)
\!\!-\!\!\left[ {\bm \sigma}_2\cdot{\bf k}_2 + \frac{ ({\bm \sigma}_2 \cdot {\bf p}^\prime_2) \,\,
{\bm \sigma}_2\cdot{\bf k}_2\,\, ({\bm \sigma}_2 \cdot{\bf p}_2)} {(E^\prime_2+m)(E_2+m)}
\right] \Bigg\},
\end{eqnarray}
where $k_i^0=E_i^\prime -E_i$ and ${\bf k}_i={\bf p}_i^\prime- {\bf p}_i$,
and the product of three Pauli matrices can be further reduced via the identity
\begin{equation}
({\bm \sigma}_i \cdot {\bf p}^\prime_i)\,\, {\bm \sigma}_i\,\, ({\bm \sigma}_i \cdot{\bf p}_i)
={\bf p}_i^\prime ({\bm \sigma}_i \cdot {\bf p}_i)+{\bf p}_i ({\bm \sigma}_i \cdot {\bf p}^\prime_i)
-{\bm \sigma}_i({\bf p}_i^\prime \cdot {\bf p}_i) +i ({\bf p}_i^\prime \times {\bf p}_i) \ .
\end{equation}
To leading order in $(p/m)^2$, the non-relativistic limits in Eqs.~(\ref{eq:j12al})--(\ref{eq:j12cl})
sum up to
\begin{eqnarray}
{\bf j}_{12}({\bf k}_1,{\bf k}_2)&=&
-i\, G^V_E(Q^2) \, \frac {f^2_{\pi NN}}{m^2_\pi} \left({\bm \tau}_1 \times {\bm \tau}_2\right)_z
\Bigg\{ \left[ {\bm \sigma}_1 \, ({\bm \sigma}_2\cdot {\bf k}_2)
\frac{f^2_\pi(k_2)}{k_2^{\, 2}+ m_\pi^2} - 1\rightleftharpoons 2 \right] \nonumber \\
&-&({\bf k}_1-{\bf k}_2)  ({\bm \sigma}_1\cdot {\bf k}_1) ({\bm \sigma}_2\cdot {\bf k}_2)
\frac{f^2_\pi(k_1)}{k_1^{\, 2}+ m_\pi^2}
\frac{f^2_\pi(k_2)}{k_2^{\, 2}+ m_\pi^2} \Bigg\} \ .
\end{eqnarray}

\section{Calculation of amplitudes}
\label{sec:amplitudes}

In this Appendix we outline the method used to compute the matrix
elements of the one-body and two-body current operators in
Eqs.~(\ref{eq:amp-pw})--(\ref{eq:amp-fsi}).  The calculation is
carried out in the Breit frame, in which the initial deuteron and
final $n$$p$ pair (with center-of-mass energy $E_k=2 \sqrt{k^2+m^2}\,$)
have velocities given respectively by
\begin{equation}
{\bf V}_i=-\frac{ {\bf q}}{2\sqrt{m_d^2+q^2/4}}\ ,\qquad
{\bf V}_f=+\frac{ {\bf q}}{2\sqrt{E_k^2+q^2/4}} \ .
\end{equation}
Here the momentum transfer ${\bf q}$ is taken along the $\hat {\bf z}$-axis.
We also define $\gamma_i$=$1/\sqrt{1-V_i^2}$ and similarly for $\gamma_f$.

The computer codes implementing the formalism discussed below have
been successfully tested by comparing, in a model calculation
which ignored boost corrections and kept only the leading terms
in the non-relativistic expansions of the one- and two-body
currents, the present results with those obtained~\cite{Schiavilla91}
with an earlier, configuration-space version of the code.

\subsection {One-body amplitude}
\label{sec:amplitude1}

Following Eq.~(\ref{eq:ampl}), we decompose the one-body amplitude into
PW and FSI amplitudes.  The PW amplitude is written as
\begin{equation}
\label {eq:amp-1bdy-pw}
{\bf A}^{\rm PW}_{1-{\rm body}}({\bf q};{\bf k},S,M_S,T,M)=
2 \int \frac {d {\bf p}}{(2 \pi)^3} \phi^\dagger_{{\bf k};SM_S,T}({\bf p}+{\bf q}/2;{\bf V}_f)\,
{\bf j}_{\perp, 1}({\bf p}_1^{\, \prime},{\bf p}_1) \, \psi_M({\bf p};{\bf V}_i) \ ,
\end{equation}
where ${\bf p}$ is the relative momentum,
${\bf p}_1^{\,\prime}$=$3\,{\bf q}/4+{\bf p}$ and
${\bf p}_1$=$-{\bf q}/4+{\bf p}$, the factor of
2 in front of the integral takes into account the identical
contribution coming from the current of nucleon
2, ${\bf j}_{\perp, 2}$, and finally $\psi_M({\bf p};{\bf V}_i)$
and $\phi_{{\bf k};SM_S,T}({\bf p}+{\bf q}/2;{\bf V}_f)$ are the momentum-space
deuteron and free $n$$p$ wave functions, boosted to the Breit
frame.  Making the change of integration variables
$[({\bf p}_\parallel+{\bf q}/2)/\gamma_f,{\bf p}_\perp] \rightarrow 
({\bf p}_\parallel,{\bf p}_\perp)$, where ${\bf p}_\parallel$
and ${\bf p}_\perp$ refer respectively to the components of
${\bf p}$ parallel and perpendicular to ${\bf q}$,
leads to the following expression for ${\bf A}^{\rm PW}_{1-{\rm body}}$
\begin{eqnarray}
\label{eq:a1pp}
{\bf A}^{\rm PW}_{1-{\rm body}}({\bf q};{\bf k},S,M_S,T,M)=
 2\,\gamma_f\int\frac {d {\bf p}}{(2 \pi)^3}
\phi^\dagger_{{\bf k};SM_S,T}({\bf p};0)\, B^\dagger({\bf p},{\bf V}_f)\,
{\bf j}_{\perp, 1}(\overline{{\bf p}}_1^{\, \prime},\overline{{\bf p}}_1) \nonumber \\
B({\bf p},{\bf V}_i)\, \psi_M[(\gamma_f{\bf p}_\parallel-{\bf q}/2)/\gamma_i,{\bf p}_\perp]
\end{eqnarray}
where $\overline{{\bf p}}_1^{\, \prime}$=$\gamma_f{\bf p}_\parallel+
{\bf q}/4+{\bf p}_\perp$ and $\overline{{\bf p}}_1$=$\gamma_f{\bf p}_\parallel-
3\,{\bf q}/4+{\bf p}_\perp$, and the boost operators $B({\bf p},{\bf V}_i)$
and $B({\bf p},{\bf V}_f)$ can be read off from Eq.~(\ref{eq:boost}) (note
that under the change of variables above, the Thomas precession term remains
unchanged, since both ${\bf V}_i$ and ${\bf V}_f$ are along ${\bf q}$).

It is convenient to expand the free $n$$p$ wave function in partial waves~\cite{Schiavilla04}:
\begin{equation}
\label{eq:partial}
 \phi_{{\bf k};SM_S,T}({\bf p};0)= \sqrt{2}\, (2\pi)^3\, \frac{\delta(k-p)}{kp}
 \sum_{LJM_J} \epsilon_{LST} \, \left[Z_{LSM_S}^{JM_J}(\hat{\bf k})\right]^*\, 
 {\cal Y}_{LSJ}^{M_J}(\hat{\bf p})\, \chi^T_0  \ ,
\end{equation}
where $\epsilon_{LST}=[1-(-)^{L+S+T}]/2$,
\begin{equation}
 Z_{LSM_S}^{JM_J}(\hat{\bf k})=\sum_{M_L} 
 \langle LM_L,SM_S|JM_J\rangle Y_{LM_L}(\hat{\bf k}) ,
\end{equation}
and ${\cal Y}_{LSJ}^{M_J}$ are standard spin-angle functions.
Inserting this expansion in Eq.~(\ref{eq:a1pp}) gives
\begin{equation}
\label{eq:jpl}
{\bf A}^{\rm PW}_{1-{\rm body}}({\bf q};{\bf k},S,M_S,T,M)=
\sum_{LJM_J}\epsilon_{LST}\, Z_{LSM_S}^{JM_J}(\hat{\bf k})\, J_{LJM_J;M}^{ST}({\bf q},k)\ ,
\end{equation}
where
\begin{eqnarray}
 J_{LJM_J;M}^{ST}({\bf q},k)= 2\, \sqrt{2}\, \gamma_f \int d\Omega_{\bf p}\,\,
 \chi^{T\, \dagger}_0\, {\cal Y}_{LSJ}^{M_J\, \dagger}(\hat{\bf p})
 \, B^\dagger({\bf p},{\bf V}_f)\,
{\bf j}_{\perp, 1}(\overline{{\bf p}}_1^{\, \prime},\overline{{\bf p}}_1) \nonumber \\
B({\bf p},{\bf V}_i)\, \psi_M[(\gamma_f{\bf p}_\parallel-{\bf q}/2)/\gamma_i,{\bf p}_\perp]
\end{eqnarray}
and the magnitude of the relative momentum is fixed
by the $\delta$-function in Eq.~(\ref{eq:partial})
to be $|{\bf p}|$=$k$ (note that ${\bf p}$ enters in the
arguments of the boost and current operators and deuteron
wave function).  For an assigned set of quantum numbers $LJM_J;M$ and
$ST$, the function $J({\bf q},k)$ is calculated efficiently by
standard Gaussian integrations over the $\hat{\bf p}$-directions.

In order to evaluate the FSI amplitude, we first introduce in Eq.~(\ref{eq:amp-fsi})
the partial wave expansions for ${\bf A}^{\rm PW}_{1-{\rm body}}$,
Eq.~(\ref{eq:jpl}), and for the $T$-matrix~\cite{Schiavilla04},
\begin{equation}
\label{eq:tmlp}
 T^{ST}_{M_S,M_S^\prime}({\bf k},{\bf k}^\prime)=
 2\, (4\pi)^2\sum_{JM_JLL^\prime} i^{L^\prime-L}
 \epsilon_{LST}\,\epsilon_{L^\prime ST}\,Z_{LSM_S}^{JM_J}(\hat{\bf k})\,
 \left[Z_{L^\prime SM_S^\prime}^{JM_J}(\hat{\bf k}^\prime)\right]^*
 T^{STJ}_{L L^\prime}(k,k^\prime) \ ,
\end{equation}
and then carry out the integrations over the $\hat{\bf k}^{\,\prime}$
solid angle to obtain
\begin{equation}
\label{eq:jplf}
{\bf A}^{\rm FSI}_{1-{\rm body}}({\bf q};{\bf k},S,M_S,T,M)=
\sum_{LJM_J}\epsilon_{LST}\, Z_{LSM_S}^{JM_J}(\hat{\bf k}) 
J_{LJM_J;M}^{ST}({\bf q},k;{\rm FSI})\ ,
\end{equation}
where we have defined
\begin{equation}
\label{eq:kpl}
J_{LJM_J;M}^{ST}({\bf q},k;{\rm FSI})=\sum_{L^\prime}
i^{L^\prime-L}\, \epsilon_{L^\prime ST}\, 
\left[ \frac{2}{\pi} \int_0^\infty dk^\prime \, k^{\prime\, 2}\, 
\frac{T^{STJ}_{L L^\prime}(k,k^\prime)}
{E_k -E_{k^\prime} + i\epsilon}
 J_{L^\prime JM_J;M}^{ST}({\bf q},k^\prime)\right]\ .
\end{equation}
In deriving the equations above, use has been made of the
following relation:
\begin{equation}
\sum_{M_S}\int d\Omega_{\bf k}\, 
\left[Z_{LSM_S}^{JM_J}(\hat{\bf k})\right]^*
\, Z_{L^\prime SM_S}^{J^\prime M_J^\prime }(\hat{\bf k})=
\delta_{JJ^\prime} \delta_{M_JM_J^\prime} \delta_{LL^\prime}\ ,
\end{equation}
while a standard subtraction technique~\cite{Gloeckle83} is employed to
perform the principal value integration implicit in Eq.~(\ref{eq:kpl}).

In the partial wave expansions of the amplitudes, fully converged results for 
${\bf A}_{\rm 1-body}^{\rm PW}$ and ${\bf A}_{\rm 1-body}^{\rm FSI}$
are obtained, at the low center-of-mass energy of the final $n\!p$
pair of interest here (1.5 MeV), when all contributions
with total angular momentum $J \le 3$ are retained in the sum over channels.

\subsection {Two-body amplitude}
\label{sec:amplitude2}

In this case, after rescaling the ${\bf p}^\prime$ relative momentum
as $({\bf p}^\prime_\parallel/\gamma_f,{\bf p}^\prime_\perp)
 \rightarrow ({\bf p}^\prime_\parallel,{\bf p}^\prime_\perp)$
in the integral, the PW amplitude reads
\begin{eqnarray}
\label{eq:a2pp}
{\bf A}^{\rm PW}_{2-{\rm body}}({\bf q};{\bf k},S,M_S,T,M)&=&
 \gamma_f\int\frac{d {\bf p}^\prime}{(2 \pi)^3}\frac{d {\bf p}}{(2 \pi)^3}
\phi^\dagger_{{\bf k};SM_S,T}({\bf p}^\prime ;0)\, B^\dagger({\bf p}^\prime,{\bf V}_f)
\nonumber \\
& &{\bf j}_{\perp, 12}(\overline{{\bf p}}_1^{\, \prime},
\overline{{\bf p}}_2^{\, \prime},\overline{{\bf p}}_1,\overline{{\bf p}}_2)
B({\bf p},{\bf V}_i)\, \psi_M({\bf p}_\parallel/\gamma_i,{\bf p}_\perp) \ ,
\end{eqnarray}
where $\overline{{\bf p}}_1^{\, \prime}$=${\bf q}/4+
\gamma_f{\bf p}^\prime_\parallel
+{\bf p}^\prime_\perp$, $\overline{{\bf p}}_2^{\, \prime}$=
${\bf q}/4-\gamma_f{\bf p}^\prime_\parallel-{\bf p}^\prime_\perp$
and $\overline{{\bf p}}_1$=$-{\bf q}/4+{\bf p}$,
$\overline{{\bf p}}_2$=$-{\bf q}/4-{\bf p}$.  Rather
than expanding the free $n$$p$ state in partial waves,
we carry out the ${\bf p}^\prime$ integration by
inserting into the equation above the plane waves
of Eq.~(\ref{eq:pw}), and obtain
\begin{eqnarray}
\label{eq:a2ppl}
{\bf A}^{\rm PW}_{2-{\rm body}}({\bf q};{\bf k},S,M_S,T,M)&=&
\sqrt{2}\,\gamma_f\int \frac{d {\bf p}}{(2 \pi)^3}
\chi^{S,T\, \dagger}_{M_S,0} \, B^\dagger({\bf k},{\bf V}_f)
\nonumber \\
& &{\bf j}_{\perp, 12}(\overline{{\bf p}}_1^{\, \prime},
\overline{{\bf p}}_2^{\, \prime},\overline{{\bf p}}_1,\overline{{\bf p}}_2)
B({\bf p},{\bf V}_i)\, \psi_M({\bf p}_\parallel/\gamma_i,{\bf p}_\perp) \ ,
\end{eqnarray}
where in the momenta $\overline{{\bf p}}_1^{\, \prime}$ and
$\overline{{\bf p}}_2^{\, \prime}$ the parallel and perpendicular
components of the relative momentum ${\bf p}^\prime$ are replaced
by those corresponding to ${\bf k}$.  The three-dimensional integrations
in Eq.~(\ref{eq:a2ppl}) are done by Gaussian quadratures.

The amplitudes ${\bf A}^{\rm FSI}_{2-{\rm body}}$ are calculated
from Eq.~(\ref{eq:amp-fsi}) by direct integration over ${\bf k}^\prime$.
To this end, we first reconstruct, from the channel solutions
$T^{STJ}_{L L^\prime}(k,k^{\,\prime})$, the full $T$-matrix in
Eq.~(\ref{eq:tmlp}), by including contributions with total
angular momentum up to $J$=3, and then use cubic-spline
techniques to interpolate the ${\bf A}^{\rm PW}_{2-{\rm body}}$,
previously tabulated on a sufficiently coarse grid,
at the ${\bf k}^\prime$ values relevant for integration. 

\newpage
%
%%%%%%%%%%%%%%%%%%%%%%%%%%%%%%%%%%%%%%%%%%%%%%%%%%%%
%     FIGURES
%%%%%%%%%%%%%%%%%%%%%%%%%%%%%%%%%%%%%%%%%%%%%%%%%%%%
%
%
\end{document}